**Implantation of labelled single nitrogen vacancy centers in diamond using $^{15}$N**

J. R. Rabeau*, P. Reichart

*School of Physics, Microanalytical Research Centre, The University of Melbourne, Parkville, Victoria 3010, Australia.*

G. Tamanyan, D. N. Jamieson, S. Prawer

*Centre for Quantum Computer Technology, School of Physics, The University of Melbourne, Parkville, Victoria 3010, Australia.*

F. Jelezko, T. Gaebel, I. Popa, M. Domhan, J. Wrachtrup

*3. Physikalisches Institut, Universität Stuttgart, Pfaffenwaldring 57, 70569 Stuttgart, Germany.*

Abstract

Nitrogen-vacancy (NV$^-$) color centers in diamond were created by implantation of 7 keV $^{15}$N (I = ½) ions into type IIa diamond. Optically detected magnetic resonance was employed to measure the hyperfine coupling of the NV$^-$ centers. The hyperfine spectrum from $^{15}$NV$^-$ arising from implanted $^{15}$N can be distinguished from $^{14}$NV$^-$ centers created by native $^{14}$N (I = 1) sites. Analysis indicates 1 in 40 implanted $^{15}$N atoms give rise to an optically observable $^{15}$NV$^-$ center. This report ultimately demonstrates a mechanism by which the yield of NV$^-$ center formation by nitrogen implantation can be measured.

* corresponding author email: jrabeau@unimelb.edu.au

The controlled fabrication of single centers in solid state systems is an important step in the development of semiconductor based quantum devices.[1] In particular, the nitrogen vacancy (NV⁻) center in diamond is a candidate for deterministic single photon generation[2] and a promising candidate for solid-state spin-based quantum computing using diamond.[3] Several proof-of-principle experiments including single spin readout[4], spin coherence lifetime measurements[5] and two qubit quantum gate operations[6] have been demonstrated recently. The latter application in particular demands a scalable single NV⁻ fabrication strategy on a nanometer scale while maintaining long coherence times.[7] One potential methodology to meet this requirement is a "top down" fabrication approach employing ion implantation.[8] The creation of NV⁻ centers in diamond has been accomplished previously by two alternate methods: (A) Ion implantation can be used to create vacancies in nitrogen-rich (native) type Ib diamond. Annealing above 600°C initiates migration of the vacancies to the native substitutional nitrogen sites to form NV⁻ centers. Implantation of gallium, helium, carbon and other ions as well as electrons have been employed to produce vacancies for NV⁻ formation with this method. It is important to emphasize that the nitrogen is already distributed randomly in the diamond crystal, and therefore only limited control over NV⁻ center formation can be achieved using this method. (B) Implantation of N ions into type IIa diamond that contains low concentrations of native N followed by annealing above 600°C also forms NV⁻ centers. In this case, the assumption is that the implanted N ions take up substitutional sites in the diamond lattice, and the vacancies created upon impact migrate to those sites forming color centers close to the site of the original ion impact. However, this has never been explicitly confirmed. This latter method may provide greater spatial control of the implantation process, which will benefit controlled fabrication of arrays of single NV⁻ centers, and the low N-content of the starting material is essential for long coherence times of the spin states of the NV⁻ center.[7]

Meijer *et al.*[9] recently implanted 2 MeV $^{14}$N ions into diamond to form NV⁻ centers. They estimated that 0.5 - 1 NV⁻ centers were produced per implanted N ion. This work however did not show whether the NV⁻ center was formed from implanted or native N's. For the case of 2 MeV $^{14}$N implants, Monte-Carlo simulations (SRIM, v.2003.26, displacement energy 55 eV,[10] density 3.5 g/cm³)[11] predict about 200 vacancies in total produced per ion along the ion track (<100 nm diameter) over a range of 1.1 µm. Even with the stated low native nitrogen content of less than 0.1 ppm there would be up to 100 *native* $^{14}$N atoms in the volume of the ion track compared to only one *implanted* $^{14}$N in the same volume. Also, there are up to 30 native $^{14}$N atoms in the nominal 150 nm diameter spherical volume surrounding the end of range where the Bragg peak in the stopping power creates the maximum concentration of vacancies. Upon annealing, it is impossible to say where the NV center originates from and, even in high purity diamond, there exists a significant possibility that the NV⁻ centers will be formed by the native $^{14}$N atoms and not by the implanted N ion. The purpose of this work is to confirm and differentiate Method A from Method B, and ultimately provide a technique for measuring the yield of NV⁻ formation.

Owing to the different nuclear spin, the hyperfine spectrum of NV⁻ centers created by the implantation of $^{15}$N (I=½) ions into type IIa diamond can be distinguished from centers created from native $^{14}$N (I = 1). This is because the hyperfine coupling in the $^3$A ground state of the $^{15}$NV⁻ center in diamond is different from the $^{14}$NV⁻ center, and this can be measured using optically detected magnetic resonance (ODMR). Therefore, any $^{15}$NV⁻ centers identified within the diamond can be attributed with a high probability to the implanted $^{15}$N ion (the natural abundance of $^{15}$N is 0.37%). Bulk electron spin resonance (ESR) measurements have previously been made which show the contrast in hyperfine coupling by directly comparing the $^{15}$NV$^-$ center with the $^{14}$NV$^-$ center[12] and ODMR has been employed previously to characterise the single $^{14}$NV$^-$ colour centers.[9] We extend this work by employing ODMR to characterise the distinct hyperfine coupling of the $^{15}$NV$^-$ center created by $^{15}$N implantation.

Type IIa diamonds (Sumitomo, <0.1 ppm native N) were irradiated with a 2 mm diameter beam to a fluence of ~$2 \times 10^9$ cm$^{-2}$ with 14 keV $^{15}$N$_2^+$ ions or $^{14}$N$_2^+$ ions and annealed for 1 hour in an Ar ambient furnace at 900 ºC. After annealing, the diamond was cleaned in a boiling acid mixture (1:1:1 H$_2$SO$_4$ : HClO$_4$ : HNO$_3$), solvents and deionized water. Scanning fluorescence microscopy was performed with a 532 nm excitation laser under continuous wave operation. The NV$^-$ center gives rise to a broad fluorescence spectrum at room temperature, with a zero-phonon line at 637 nm and intense phonon side bands extending to ~720 nm.[13] Optical filters were employed to block the pump beam and selectively sample a large portion of the phonon side bands from the NV$^-$ fluorescence (532 nm notch filter and a 665 nm long pass filter). The long-pass filter also blocked the first and second order Raman lines. At the selected ion implantation fluence (~$10^9$ $^{15}$N ions/cm$^2$), the average separation between N impact sites is of the same order as the spatial resolution of the fluoresecence measurements (i.e. about 0.3 μm). Unimplanted samples showed no fluorescence from NV$^-$ centers. Implanted and annealed samples showed many NV$^-$ centers as shown in Figure 1, which is a 2D confocal fluorescence image of a sample implanted with $^{15}$N$_2^+$. Each bright spot corresponds to an optically active NV$^-$ site, which, at the given dose, could potentially correspond to either 1 or 2 NV$^-$ centers.

Twenty fluorescing sites were selected for further analysis. Photon antibunching[14] measurements confirmed that each of these twenty spots were single NV⁻ centers. The sample was then placed within a micro-coil induced tunable microwave field and the hyperfine substructure of the ³A state probed using the ODMR technique. This technique employed the same excitation and fluorescence-collection as discussed above while focused on one single NV⁻ center, and a weak microwave field swept over the range of interest (around 2.88 GHz).

The spin Hamiltonian of the NV⁻ center is $H=\mathbf{SDS}+g\beta \mathbf{SH}_0+\mathbf{SAI}+g_n\beta_n \mathbf{IH}_0$, where $\mathbf{D}$ and $\mathbf{A}$ are the fine structure and hyperfine splitting (hfs) tensors, $g$, $g_n$ and $\beta$, $\beta_{e,n}$ are the electron and nuclear g-factors and Bohr magnetons, respectively. Splitting of the $m_s= 0$ and $m_s= \pm 1$ ground state sublevels in the $^{14}$NV⁻ centre is known to be 2.88 GHz under zero B-field.[15] In our experiments the degeneracy of $m_s= \pm 1$ sublevels was lifted by applying a weak (0.02 T) magnetic field allowing to selectively address the $m_s= 0$ to $m_s= 1$ transition, thus making for a cleaner spectrum.

By sweeping the applied microwave field and simultaneously probing the optical transition with the 532 nm laser beam, it is possible to specifically observe the hyperfine structure via a change in fluorescence intensity. The microwave field was deliberately made weak in order to eliminate the line broadening existent under strong applied fields and allow the hyperfine structure to be resolved. The hyperfine splitting of the ground state triplet of the $^{14}$NV⁻ and $^{15}$NV⁻ center gave rise to distinct ODMR spectra which corresponded to the theoretically calculated energy level schemes shown in Figure 2. The inset in Figure 3 shows the triplet ODMR spectrum from the $^{14}$NV⁻ as predicted.[16]

The hyperfine coupling constant associated with the $^{15}$N nucleus has been calculated according to the procedure described previously.[9] For the axially symmetric case, the hyperfine tensor has the components $A_\parallel$ and $A_\perp$:

$$A_\parallel = a + 2b$$
$$A_\perp = a - b$$

with $a$ (isotropic term) and $b$ (anisotropic term) given by

$$a = (8\pi/3) g \beta g_n \beta_n c_{2s}^2 \eta^2 |\psi_{2s}(r_n)|^2$$

$$b = \frac{2}{5} g \beta g_n \beta_n c_{2p}^2 \eta^2 \langle 1/r_{2p}^3 \rangle$$

where $\eta^2$ is the spin density and $|\psi_{2s}(r_n)|^2$ the 2s wave function at the nucleus. We use the spin density corresponding to the case of $^{14}$N, while the coefficients corresponding to the 2s and 2p orbitals were obtained from fitting the experimental ODMR data. With $g_n$ = -0.5664 for the $^{15}$N nucleus, the calculated isotropic term is $a$ = –3.05 MHz and the anisotropic term $b$ = 0. Therefore, the hyperfine interaction with the $^{15}$N nucleus can be considered isotropic with an expected value of –3.05 MHz. This was confirmed experimentally in Figure 3 which shows a splitting of 3.1 MHz for the $^{15}$NV$^-$ center. The energy level scheme in Figure 2 indicates 3 possible transitions in the hyperfine spectrum for the $^{14}$NV$^-$ center, and 2 possible transitions for the $^{15}$NV$^-$ center. The experimental data shown in Figure 3 clearly reflects this signature with both a doublet ($^{15}$NV$^-$) and triplet ($^{14}$NV$^-$) spectrum.

The ODMR spectra from NV$^-$ centers in the $^{15}$N implanted diamond therefore provide conclusive evidence that the implanted nitrogen becomes substitutional in the diamond crystal and is ultimately responsible for the formation of the $^{15}$NV$^-$ center.

Of the twenty bright centers examined, samples implanted with $^{14}$N were measured to have only the $^{14}$NV$^-$ ODMR spectrum, whereas ODMR spectra from samples implanted with $^{15}$N gave rise only to the $^{15}$NV$^-$ spectrum.

The ability to differentiate between a $^{15}$NV$^-$ and $^{14}$NV$^-$ center allows for determination of the yield of optically detectable NV$^-$ centers per implanted N ion. The number of single NV$^-$ centers per unit area was compared to the fluence of the implanted $^{15}$N ions. Under the conditions employed here, a yield of approximately (2.5±0.6)% single $^{15}$NV$^-$ centers per implanted $^{15}$N was obtained. This is low in comparison to the yield of about 50% recorded by Meijer *et al.*[9.] for high energy implants. However, as discussed above, in the high energy implantation case, the ion track volume is more than three orders of magnitude larger than that of the present work, and contains up to 30 $^{14}$N atoms in the end of range peak (and even more in the total ion track volume) which are available for NV formation. It is also a possibility which cannot be discounted that the yield is inherently higher with MeV ion implants for some other reason.

Many factors may influence the yield of observable NV$^-$ centers. Amongst these are proximity to the surface, the background doping level (which may influence the ratio of NV$^-$ to NV$^0$), the implantation and annealing temperatures (which can influence the relative proportion for the formation of NV$^-$ to other competing forms of N in diamond) and the effectiveness of defect removal (since residual defects could quench NV$^-$ luminescence). Although the technique we employed only yielded 1 NV in 40 implanted N's, the important point of this paper is that we have identified and demonstrated a contrast mechanism which enables us to measure this number. Until now it was not known what the yield for creating NV's by implanting N was, because it was not known how to differentiate implanted versus native nitrogen. With the contrast mechanism of $^{15}$N versus $^{14}$N, it is now possible to measure the yield. With this capability, we now have an essential tool which will enable us to improve the formation yield. This result is another step toward high spatial accuracy creation of the NV$^-$ center by ion implantation techniques, and ultimately arrayed NV$^-$ centers.

Acknowledgements

This work has been supported by ARO grant under contract number W911NF-05-1-0284, the Australian Research Council Discovery Project , the DFG SFB/TR21 and Landesstiftung BW Atomoptik. The authors wish to thank Rafi Kalish, Paolo Olivero, Andrew Greentree, Elizabeth Trajkov and Phil Hemmer for useful discussions.

Figure Captions

FIG 1. Room temperature confocal fluorescence image from a type IIa single crystal diamond implanted with $^{15}N_2^+$. The diamond was illuminated with a 532 nm laser through a 100X oil immersion objective lense. Fluorescence was collected through the same lens and transmitted through a 665 nm long pass filter to select only fluorescence from the $NV^-$ centre. Each bright spot is an optically active $^{15}NV^-$ site.

FIG 2. Energy level schemes for the $^{15}NV^-$ and the $^{14}NV^-$ colour centers showing the difference in hyperfine coupling in the ground state spin substructure. The schemes reflect the energy levels under a weak 0.02 T magnetic field which lifts the degeneracy of the $m_s= \pm 1$ sublevels.

FIG 3. Optically detected magnetic resonance spectra from single $^{15}NV^-$ and $^{14}NV^-$ colour centers collected under a 0.02 T magnetic field. The clear distinction between the two spectra showed that implanted ions were giving rise to the $NV^-$ center.

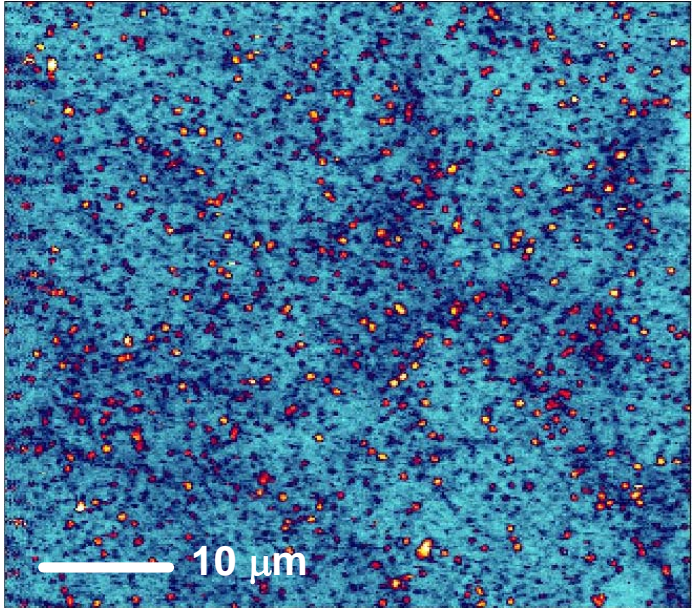

Figure 1. J. R. Rabeau *et al.*

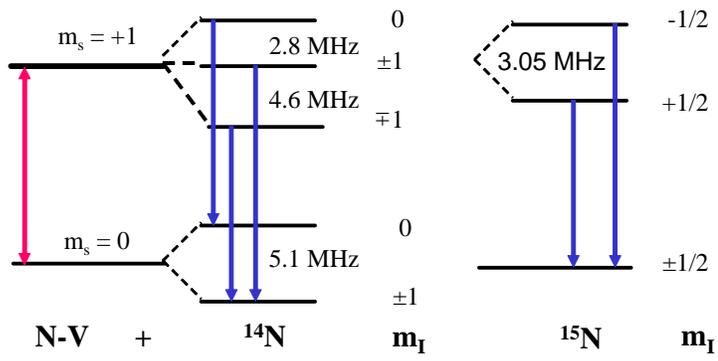

Figure 2. J. R. Rabeau *et al.*

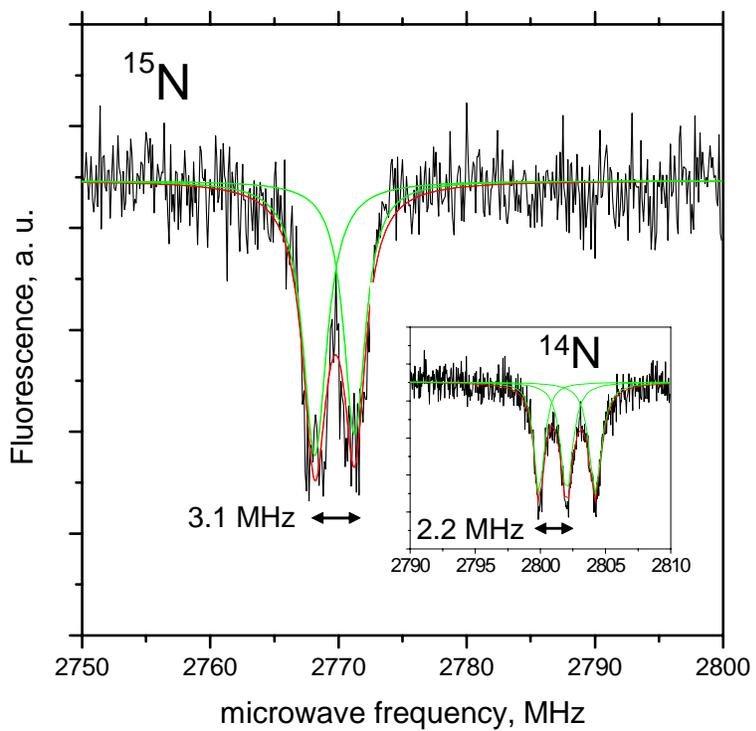

Figure 3. J. R. Rabeau *et al.*